\documentclass[sn-mathphys]{sn-jnl}
\usepackage{physics}
\usepackage{siunitx}

\begin{document}

\title[Article]{
    \textit{Ab initio} Real-Time Quantum Dynamics of Charge Carriers in
    Momentum Space
    }

\author[1]{\fnm{Zhenfa} \sur{Zheng}}
\author*[1,2,3]{\fnm{Yongliang} \sur{Shi}} \email{sylcliff@xjtu.edu.cn}
\author[4]{\fnm{Jin-jian} \sur{Zhou}}
\author[5]{\fnm{Oleg V.} \sur{Prezhdo}}
\author*[1]{\fnm{Qijing} \sur{Zheng}} \email{zqj@ustc.edu.cn}
\author*[1,6]{\fnm{Jin} \sur{Zhao}} \email{zhaojin@ustc.edu.cn}

\affil*[1]{
    \orgdiv{Department of Physics, ICQD/Hefei National Research Center for Physical
            Sciences at Microscale},
    \orgname{University of Science and Technology of China},
    \orgaddress{\city{Hefei}, \postcode{230026}, \state{Anhui}, \country{China}}
    }

\affil*[2]{
    \orgdiv{Center for Spintonics and Quantum Systerms, State Key Laboratory
            for Mechanical Behavior of Materials, School of Materials Science
            and Engineering},
    \orgname{Xi'an Jiaotong University},
    \orgaddress{\city{Xi'an}, \postcode{710049}, \state{Shanxi}, \country{China}}
    }

\affil*[3]{
    \orgdiv{China 2 State Key Laboratory of Surface Physics and Department of Physics},
    \orgname{Fudan University},
    \orgaddress{\city{Shanghai}, \postcode{200433}, \country{China}}
    }

\affil[4]{
    \orgdiv{School of Physics},
    \orgname{Beijing Institute of Technology},
    \orgaddress{\city{Beijing}, \postcode{100081}, \country{China}}
    }

\affil[5]{
    \orgdiv{Departments of Chemistry, Physics, and Astronomy},
    \orgname{University of Southern California},
    \orgaddress{\city{Los Angeles}, \postcode{90089}, \state{California}, \country{USA}}
    }

\affil*[6]{
    \orgdiv{Department of Physics and Astronomy},
    \orgname{University of Pittsburgh},
    \orgaddress{\city{Pittsburgh}, \postcode{15260}, \state{Pennsylvania}, \country{USA}}
    }

    \abstract{
      Application of the nonadiabatic molecular dynamics (NAMD) approach is
      severely limited to studying carrier dynamics in the momentum space,
      since a supercell is required to sample the phonon excitation and
      electron-phonon (\textit{e-ph}) interaction at different momenta in a
      molecular dynamics simulation.  Here, we develop an \textit{ab initio}
      approach for the real-time quantum dynamics for charge carriers in the
      momentum space (NAMD\_$\vb{k}$) by directly introducing the \textit{e-ph}
      coupling into the Hamiltonian based on the harmonic approximation. The
      NAMD\_$\vb{k}$ approach maintains the quantum zero-point energy and
      proper phonon dispersion, and includes memory effects of phonon
      excitation.  The application of NAMD\_$\vb{k}$ to the hot carrier
      dynamics in graphene reveals the phonon-specific relaxation mechanism. An
      energy threshold of \SI{0.2} {eV}, defined by two optical phonon modes
      strongly coupled to the electrons, separates the hot electron relaxation
      into fast and slow regions with the lifetimes of pico- and nano-seconds,
      respectively.   The NAMD\_$\vb{k}$ approach provides a powerful tool to
      understand real-time carrier dynamics in the momentum space for different
      materials.
      }

\maketitle

\section{Introduction}\label{sec1}

Tracking the quantum dynamics of excited charge carriers in solid materials in
multi-dimensions including time and energy domains, as well as real and
momentum spaces is fundamental to the understanding of many dynamical processes
in optoelectronics, spin- and valley-tronics, solar energy conversion, and so
on.\cite{RN371, RN354, RN251, RN473, RN225, RN515} Different time-resolved
experimental techniques, including ultrafast time- and angle-resolved
photoemission spectroscopy (TR-ARPES) with time, energy, and momentum
resolution, have been rapidly developed and applied to investigate charge
carrier dynamics in various materials.\cite{RN598, RN601, RN600} However,
without the input of \textit{ab initio} investigations, it is rather difficult
to understand the physical mechanisms behind the experimental spectra. Thus, it
is urgent to develop an \textit{ab initio} simulation approach to achieve a
state-of-the-art understanding of multi-dimensional carrier dynamics in solids.

\textit{Ab initio} approach based on perturbation theory provides useful
information to understand carrier lifetimes governed by different scattering
mechanisms.  However, the real-time dynamics information cannot be obtained
straightforwardly. In recent decades, the real-time time-dependent density
functional theory (TDDFT) based on Ehrenfest dynamics\cite{RN412, RN408, RN333,
RN312, RN414} and nonadiabatic molecular dynamics (NAMD) approaches combining
time-dependent Kohn-Sham (TDKS) theory and surface hopping have been applied to
investigate the quantum dynamics of excited charge carriers.\cite{RN515, RN222,
RN210, RN413, RN594} The electron-phonon (\textit{e-ph}) coupling, spin-orbit
coupling (SOC), and many-body electron-hole interaction have been included with
different theoretical strategies.\cite{RN497, RN494, RN109, RN542, RN570,
RN597} However, in all these methods, it is rather difficult to achieve
real-time carrier dynamics in the momentum space. In previous NAMD simulations,
the phonon excitation is described using \textit{ab initio} molecular dynamics
(AIMD) within periodic boundary conditions, and only phonons at the $\Gamma$
point are included. Thus, the electron transition from one $\vb{k}$ to another
through \textit{e-ph} is forbidden even though the electronic states can be
simulated using multi-$\vb{k}$ grids. To sample the phonon excitation and
electron-phonon interaction at different momenta, a supercell needs to be used,
so that phonons at other $\vb{q}$-points can be folded to the $\Gamma$ point.
The $\vb{q}$-grid density is determined by the size of the supercell, and due
to the computational cost, usually, only a few $\vb{q}$-points can be included
in the NAMD simulation.\cite{RN109, RN497, RN150, RN593, RN515} As a contrast,
the \textit{e-ph} scattering between different momenta often needs to be
simulated with a very dense of $\vb{k}$- and $\vb{q}$-grids, especially when
the electronic band dispersion is strong. Therefore, an \textit{ab initio}
approach to describe the real-time quantum dynamics of photoexcited carriers in
the momentum space is essential.

In this work, by introducing the \textit{e-ph} coupling elements into the
time-propagation Hamiltonian, we have extended the \textit{ab initio} NAMD
approach from the real-space (labeled as NAMD\_$\vb{r}$) to the momentum space
(NAMD\_$\vb{k}$).  Different from the previous NAMD\_$\vb{r}$ approach, where
an AIMD simulation using a large supercell is required for $\vb{q}$-grid
sampling, in NAMD\_$\vb{k}$ the $\vb{k}$ and $\vb{q}$ sampling is performed by
the calculation of \textit{e-ph} matrix elements using a unit cell. The
computational cost is significantly reduced. Moreover, the NAMD\_$\vb{k}$
approach provides a straightforward picture not only for the dynamics of
excited electrons in the momentum space, but also the time-dependent phonon
excitation of the lattice due to the \textit{e-ph} scattering.  The phonon
zero-point energy as well as the phonon dispersion are accurately represented
with memory effects.  Using this approach, we have investigated the hot carrier
dynamics in graphene. It is found that there is an energy threshold at
\SI{0.2}{eV} above the Fermi level ($E_f$). The threshold separates the
hierarchical relaxation dynamics from fast [picosecond(ps)] to slow
[nanosecond(ns)] regions. The intervalley \textit{e-ph} scattering is activated
in the fast region but strongly suppressed in the slow region.  The energy
threshold is determined by strongly coupled optical phonon modes $A_1$ and
$E_{2g}$. Our work not only reveals the phonon mode-specific energy threshold
for hot electron relaxation in graphene, but also provides a powerful tool
which can be widely applied to study excited carrier dynamics in different
solid state systems with momentum space resolution.

\section{Methodology}\label{sec2}

In the NAMD\_$\vb{r}$ approach, the charge carrier (electron or hole)
wavefunction is expanded in the basis of instantaneous adiabatic Kohn-Sham (KS)
orbitals, which are obtained by solving the KS equation at atomic configuration
$\vb{R}(t)$,
\begin{equation}\label{eq:Psi1}
  \ket{\Psi(\vb{r}; \vb{R}(t))}
  = \sum_n c_n(t) \ket{\psi_n(\vb{r}; \vb{R}(t))}.
\end{equation}
Based on the classical-path approximation (CPA), $\vb{R}(t)$ can be obtained by
AIMD.  The charge carrier wavefunction follows the time-dependent Schr\"odinger
equation (TDSE)
\begin{equation}\label{eq:TDSE}
  i\hbar\pdv{t} \ket{\Psi(\vb{r};\vb{R}(t))}
  = \hat{H}^{el}(\vb{r};\vb{R}(t)) \ket{\Psi(\vb{r};\vb{R}(t))}.
\end{equation}
Then, a set of differential equations for the coefficients $c_m(t)$ is produced
\begin{equation}
  i\hbar \dot{c}_m(t) = \sum_n c_n(t) [\epsilon_n - i\hbar d_{mn}],
\end{equation}
where $\epsilon_n$ is the energy of the adiabatic KS state, and $d_{mn}$ is the
NAC between KS states $m$ and $n$. The NAC can be written as
\begin{equation}
  d_{mn}
  = \mel{\psi_m}{\dv t}{\psi_n}
  = \frac{\mel{\psi_m}{\Delta_{\vb{R}} \hat{H}^{el}}{\psi_n}}{\epsilon_n -
    \epsilon_m} \cdot \dot{\vb{R}}.
\end{equation}
Here, $\epsilon_m$ and $\epsilon_n$ are the eigenvalues of the KS orbitals $m$
and $n$, $d_{mn}$ is the \textit{e-ph} coupling term, and $\dot{\vb{R}}$ is the
nuclear velocity. NAC is the crucial term in the NAMD\_$\vb{r}$ simulation.  It
determines not only the time-dependent coefficient evolution, but also the
hopping probability in the subsequent surface hopping step.\cite{RN210, RN222}
According to Bloch’s theory, if $\psi_m$ and $\psi_n$ have different $\vb{k}$
vectors, the NAC is zero (see more details in the Supplementary Materials). The
essential reason is that in the AIMD simulation with periodic conditions, only
phonon modes at the $\Gamma$ point are included. Therefore, the NAMD\_$\vb{r}$
approach can not efficiently simulate the carrier dynamics in the momentum
space.

The NAMD\_$\vb{k}$ approach is based on the harmonic approximation. Here,
different with eq.~(\ref{eq:Psi1}), we expand the charge carrier wavefunction
using the KS orbitals of the equilibrium atomic configuration
$\vb{R}_0$
\begin{equation}
  \ket{\Psi(\vb{r};\vb{R}(t))}
  = \sum_{n\vb{k}} c_{n\vb{k}}(t) \ket{\psi_{n\vb{k}}(\vb{r};\vb{R}_0)},
\end{equation}
where the KS orbital $\ket{\psi_{n\vb{k}}(\vb{r};\vb{R}_0)}$ with band index
$n$ and momentum $\vb{k}$ is the eigenstate of the equilibrium configuration
$\vb{R}_0$. Naturally, the charge carrier Hamiltonian is divided into two
parts
\begin{equation}\label{eq:Ham}
  \hat{H}^{el}(\vb{r};\vb{R}(t))
  = \hat{H}^0(\vb{r};\vb{R}_0) + \Delta V(\vb{r};\vb{R}(t)),
\end{equation}
where $\Delta V$ is the variation of the potential induced by nuclear
displacements $\Delta\vb{R}(t)=\vb{R}(t)-\vb{R}_0$. Combining the above
equations, we get a new coefficient evolution equation
\begin{equation}\label{eq:TDCoef}
  i\hbar\dv{t}c_{m\vb{k'}}(t) = \sum_{n\vb{k}}
    (H^0_{m\vb{k}',n\vb{k}} + H^{ep}_{m\vb{k}',n\vb{k}}) c_{n\vb{k}}(t).
\end{equation}
Here,
\begin{equation}\label{eq:H0}
  H^0_{m\vb{k}',n\vb{k}} = \mel{\psi_{m\vb{k}'}}{\hat{H}^0}{\psi_{n\vb{k}}}
           = \epsilon_{n\vb{k}}\delta_{mn,\vb{k}'\vb{k}},
\end{equation}
is the diagonal KS energy matrix and
\begin{equation}\label{eq:Hep0}
  H^{ep}_{m\vb{k}',n\vb{k}} = \mel{\psi_{m\vb{k}'}}{\Delta V}{\psi_{n\vb{k}}},
\end{equation}
is the \textit{e-ph} coupling Hamiltonian. $m$ and $n$ are the notation of
the KS orbitals, and $\vb{k}$ and $\vb{k}'$ are the notation of the momentum.

Transformed into the momentum space, the \textit{e-ph} term can be rewritten
as
\begin{equation}
  \begin{aligned}\label{eq:EPH}
    H_{{m\vb{k}'},{n\vb{k}}}^{ep}
    &= \frac{1}{\sqrt{N_p}} \sum_{\vb{q}\nu}
       \mel{u_{m\vb{k}'}}{\Delta_{\vb{q}\nu}v(\vb{r};\vb{R}_0)}{u_{n\vb{k}}}_{uc}
       \delta_{\vb{q},\vb{k}'\text{-}\vb{k}} Q_{\vb{q}\nu}(t) / l_{\vb{q}\nu} \\
    &= \frac{1}{\sqrt{N_p}} \sum_{\nu}
       \eval{g_{mn\nu}(\vb{k},\vb{q}) Q_{\vb{q}\nu}(t) /
       l_{\vb{q}\nu}}_{\vb{q}=\vb{k}'\text{-}\vb{k}},
  \end{aligned}
\end{equation}
where $N_p$ is the number of unit cells according to the Born-von
K\'{a}m\'{a}nn boundary conditions, $Q_{\vb{q}\nu}(t)$ is the normal mode
coordinate of the corresponding vibration mode of phonon with momentum $\vb{q}$
in the branch $\nu$, $l_{\vb{q}\nu}$ is the “zero-point” displacement
amplitude, and $g_{mn\nu}(\vb{k},\vb{q})$ is the \textit{e-ph} matrix element.
In this way, the NAC in the NAMD\_$\vb{r}$ approach is replaced by the
\textit{e-ph} coupling Hamiltonian in eq.~(\ref{eq:EPH}), which naturally
includes the coupling between electronic states with different momenta $\vb{k}$
and the scattering with phonons at different momenta $\vb{q}$.

In the NAMD\_$\vb{k}$ method, to get the real-time carrier dynamics, the
\textit{e-ph} coupling matrix element $g_{mn\nu}(\vb{k},\vb{q})$ and the
time-dependent normal mode coordinate $Q_{\vb{q}\nu}(t)$ are
required. $g_{mn\nu}(\vb{k},\vb{q})$ can be calculated by the DFPT method using
the primitive cell\cite{RN164} or finite difference method with non-diagonal
supercells.\cite{RN596} $Q_{\vb{q}\nu}(t)$ can be obtained using different
methods. For example, it can be expressed in terms of phonon populations as
\begin{equation}
  Q_{\vb{q}\nu}(t)
  = l_{\vb{q}\nu} \sqrt{n_{\vb{q}\nu}+\frac{1}{2}}
   ( e^{-i\omega_{\vb{q}\nu}t} + e^{i\omega_{\text{-}\vb{q}\nu}t} ),
\end{equation}
where the initial population of phonons at $t=0$ ($t_0$) is given by the
Bose-Einstein distribution $n_{\vb{q}\nu} = \frac{1}{e^{\hbar
\omega_{\vb{q}\nu} / k_B T}-1} $. It can also be obtained from the molecular
dynamics simulation using the normal mode decomposition method.\cite{RN572}
Finally, the fewest switches surface hopping (FSSH) is applied to include the
stochastic factor of the carrier dynamics. More details can be found in the
Supplementary Materials.

\section{Results and Discussion}\label{sec3}

To verify the validity of the NAMD\_$\vb{k}$ method, we choose graphene as a
prototypical system and simulate the hot electron relaxation process, which has
been investigated extensively.\cite{RN492, RN480, RN226, RN233, RN241, RN238,
RN231, RN237, RN489, RN488, RN567} The band structure of graphene has six Dirac
cones near the $E_f$ at $K$ and $K'$ points in the first Brillouin zone (BZ),
which is also known as six valleys. As discussed in the Supplementary Materials,
the NAMD\_$\vb{k}$ approach using the $9\times9\times1$ $\vb{k}$-grid can
successfully reproduce the NAMD\_$\vb{r}$ results using the $9\times9\times1$
supercell.

\begin{figure}[ht]
\centering
\includegraphics[width=1.0\textwidth]{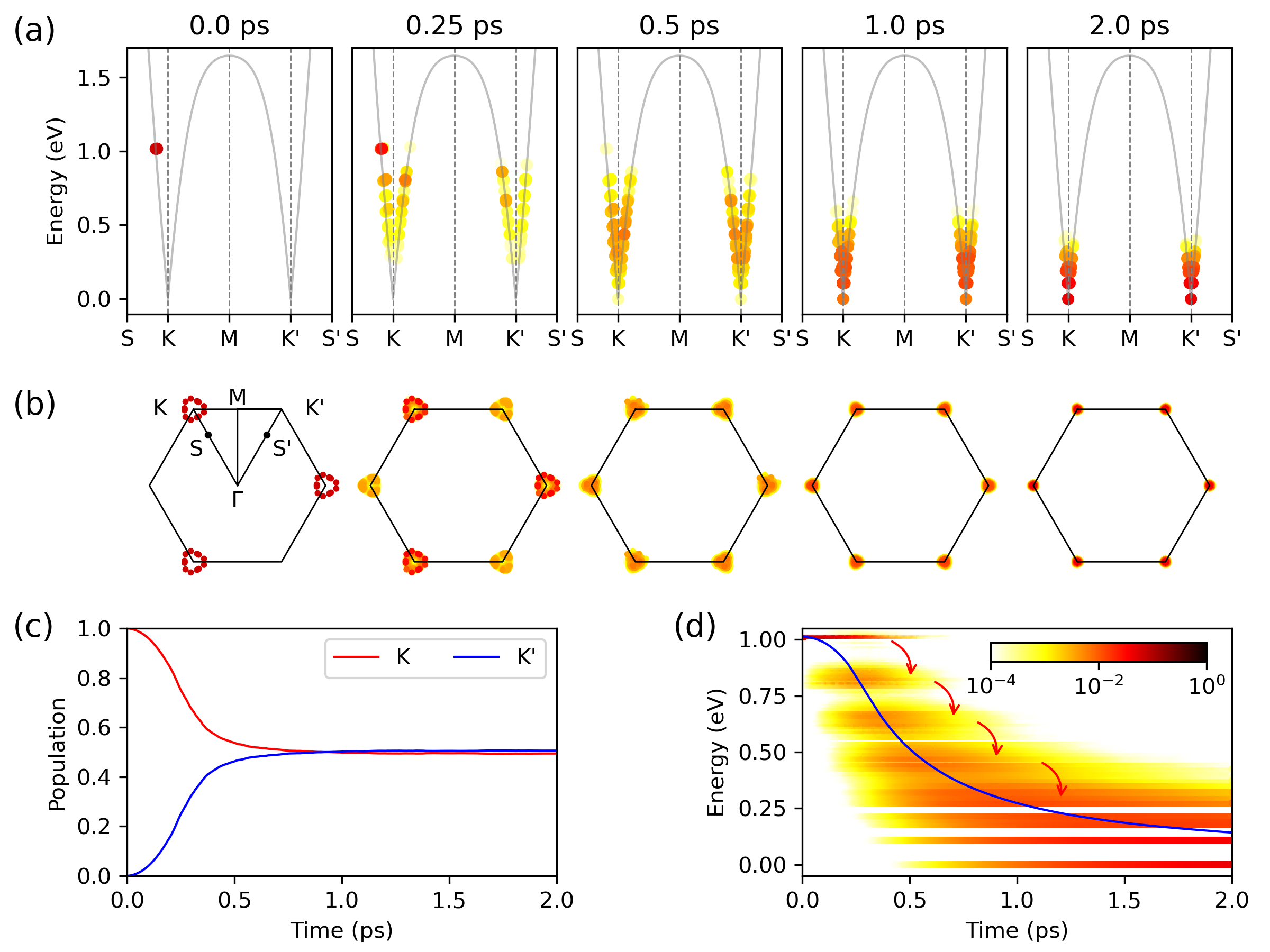}
\caption{
  Hot electron relaxation in graphene with $E_\text{ini} = \SI{1.0}{eV}$ using
  $150\times150\times1$ $\vb{k}$-grid. (a-b) Snapshots of hot electron
  distribution in energy and momentum space at
  \SIlist[list-units=single]{0;0.25;0.5;1.0;2.0}{ps}. The color dots indicate
  the electron population in different states. (c) Time-dependent electron
  population in the $K$ and $K'$ valleys.(d) Hot electron relaxation in the
  energy domain. The color strips indicate the electron population in different
  energy states, and the blue line represents the averaged electron energy.
  The energy reference is $E_f$.
  }\label{fig1}
\end{figure}

The hot electron relaxation may involve intervalley and intravalley
\textit{e-ph} scattering. We first study dynamics with a single electron
initially excited at \SI{1.0}{eV} above the $E_f$ ($E_\text{ini} =
\SI{1.0}{eV}$) in the $K$ valley. To get statistics on the quantum behavior of
the excited electron, we randomly set the initially excited electron at 30
different $\vb{k}$-points in the $K$ valley at \SI{1.0}{eV} above the $E_f$.
For each $\vb{k}$-points we sample $2\times10^4$ trajectories.  Using
$150\times150\times1$ $\vb{k}$-grid is found to be able to achieve well
converged results (the $\vb{k}$-grid convergence details are presented in the
Supplementary Materials) . Figure~\ref{fig1} (a) shows five snapshots of the
hot electron population in the band structure over \SI{2}{ps}, and
Figure~\ref{fig1} (b) gives the corresponding hot electron distribution in the
first BZ. It can be seen that although the hot electron is initially excited in
the $K$ valley, the $K$-$K'$ intervalley scattering almost immediately starts.
Figure~\ref{fig1} (c) presents the time-dependent electron population in the
$K$ and $K'$ valleys. The valley lifetime ($\tau_{K}$), which is defined as the
timescale when the equilibrium between $K$ and $K'$ is reached, is around
\SI{0.4}{ps}. The intervalley scattering suggests that the hot electron couples
with phonons with large momentum.  Figure~\ref{fig1} (d) shows the hot electron
energy relaxation. The color bar indicates the electron population, and the
blue line represents the averaged energy. It can be seen that there is an
energy threshold for hot electron relaxation, located at around \SI{0.2}{eV}
above the $E_f$. Above the threshold energy, the relaxation is a relatively
fast process, which corresponds to energy relaxation from \SI{1.0}{eV} to
around \SI{0.2}{eV} within \SI{2}{ps}.  Using a Gaussian function, the lifetime
for this fast energy relaxation ($\tau_E$) can be estimated to be
\SI{0.56}{ps}. Furthermore, a quantized character with an energy difference of
around \SI{0.2}{eV} as indicated in Figure~\ref{fig1} (d). Following the fast
process, there happens a much slower relaxation process from \SI{0.2}{eV} to
the Dirac point. The timescale of the slow process is difficult to be estimated
with a \SI{2}{ps} simulation.

\begin{figure}[ht]
\centering
\includegraphics[width=1.0\textwidth]{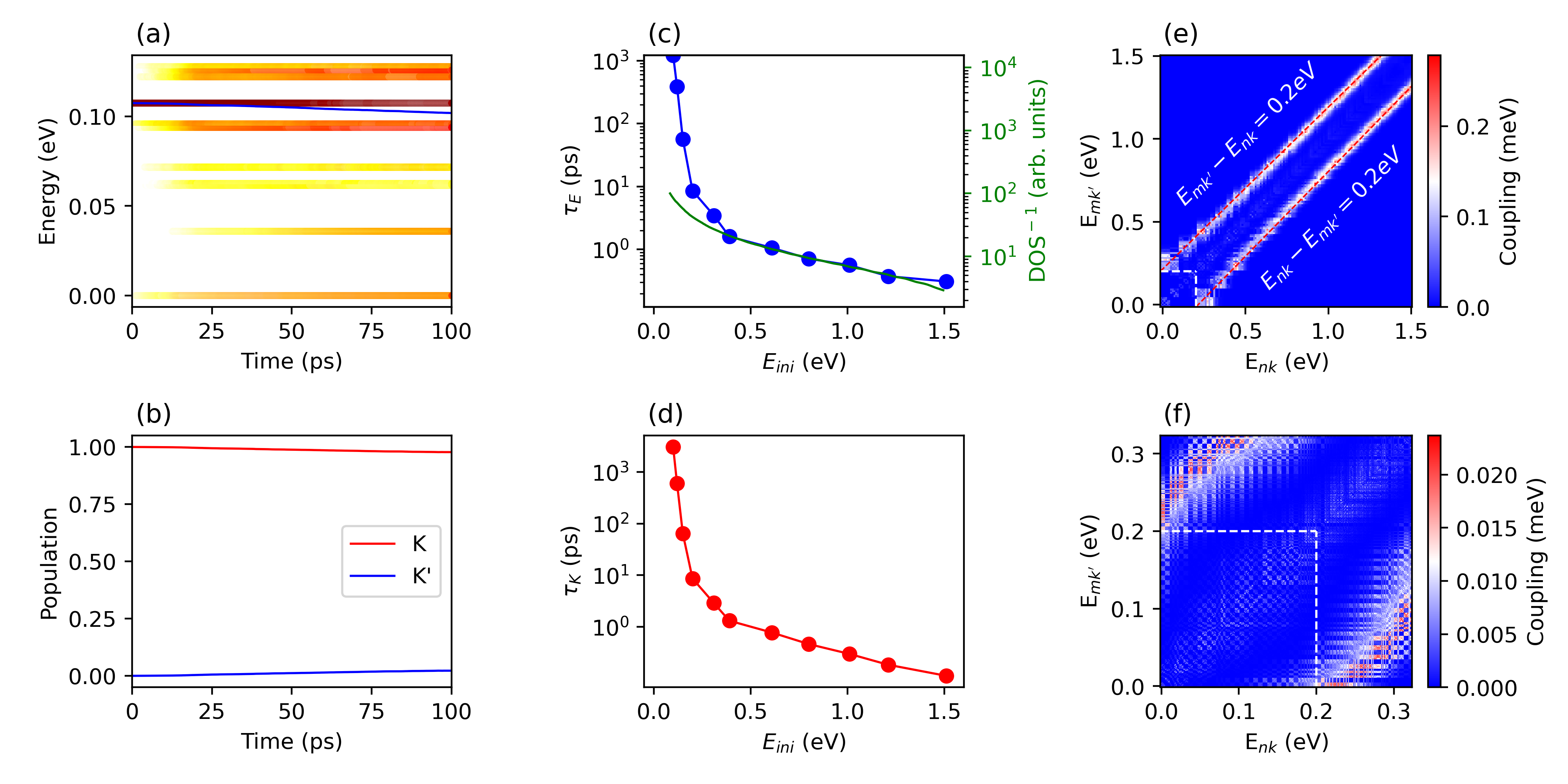}
\caption{
  Hot electron relaxation in graphene with different $E_\text{ini}$. (a) Hot
  electron relaxation in energy domain with $E_\text{ini} = \SI{0.1}{eV}$. The
  color strips indicate the electron distribution in different energy states,
  and the blue line represents the averaged electron energy. The energy
  reference is $E_f$. (b) Time-dependent electron population in the $K$ and
  $K'$ valleys with $E_\text{ini} = \SI{0.1}{eV}$. (c) The dependence of
  $\tau_E$ on $E_\text{ini}$.  When $E_\text{ini}> \SI{0.2}{eV}$,
  $150\times150\times1$ $\vb{k}$-grid is used. There are fast and slow
  relaxation processes, and $\tau_{\scriptscriptstyle E}^\text{fast}$ is
  plotted. When $E_\text{ini}< \SI{0.2}{eV}$, $450\times450\times1$
  $\vb{k}$-grid is used. There is only a slow relaxation process, and
  $\tau_{\scriptscriptstyle E}^\text{slow}$ is plotted. The green line
  represents the reciprocal of density of states (DOS$^{-1}$) at different
  energies. $\tau_{\scriptscriptstyle E}^\text{fast}$ is proportional to
  DOS$^{-1}$. The energy reference is $E_f$. (d) The dependence of $\tau_K$ on
  $E_\text{ini}$. (e-f) \textit{E-ph} coupling $H_{{m\vb{k}'}, n\vb{k}}^{ep} =
  \mel{\psi_{m\vb{k}'}}{\Delta V}{\psi_{n\vb{k}}}$ between states
  $\psi_{m\vb{k}'}$ and $\psi_{n\vb{k}}$ plotted with different $E_{m\vb{k}'}$
  and $E_{n\vb{k}}$ scales.
  }\label{fig2}
\end{figure}

To understand further the slow relaxation process close to the $E_f$, we
perform a \SI{100}{ps} NAMD\_$\vb{k}$ simulation for hot electron relaxation
with $E_\text{ini} = \SI{0.1}{eV}$. In this case, since the density of states
(DOS) become smaller when the energy is close to the $E_f$, a more dense of
$\vb{k}$-grid is required, and we use $450\times450\times1$ $\vb{k}$-grid.  The
energy and valley dynamics are shown in Figure~\ref{fig2} (a-b). Both the
energy and valley relaxation become much slower, and there is no longer a fast
process in energy relaxation. $\tau_E$ and $\tau_K$ are estimated to be
\SIlist[list-units=single]{1.2;3.0}{ns}, respectively. Such a relaxation
behavior is completely different from $E_\text{ini} = \SI{1.0}{eV}$. We further
study the hot electron relaxation dynamics with different $E_\text{ini}$ from
\SIrange[range-units=single]{0.1}{1.5}{eV}. As shown in Figure~\ref{fig2} (c),
it is found that $E_\text{ini} = \SI{0.2}{eV}$ is a critical point for
different relaxation behaviors.  If $E_\text{ini}> \SI{0.2}{eV}$, there will be
both a fast and a slow relaxation process.  The $\tau_E$ for the fast process
($\tau_{\scriptscriptstyle E}^\text{fast}$) ranges from
\SIrange[range-units=single]{0.3}{3.0}{ps}, inversely proportional to the DOS.
When $E_\text{ini}< \SI{0.2}{eV}$, there is only the slow process and
$\tau_{\scriptscriptstyle E}^\text{slow}$ dramatically increases by 1-3 orders
of magnitude. The correlation between $\tau_K$ and $E_\text{ini}$ is shown in
Figure~\ref{fig2} (d), demonstrating a very similar trend with $\tau_E$. When
$E_\text{ini}> \SI{0.2}{eV}$, the equilibrium between the $K$ and $K'$ valleys
can be reached on a ps timescale due to frequent intervalley \textit{e-ph}
scattering. Whereas, when $E_\text{ini}< \SI{0.2}{eV}$, the intervalley
\textit{e-ph} scattering becomes rare, and $\tau_K$ reaches a ns timescale. The
slow energy relaxation process is dominated by the intravalley scattering.

In the NAMD\_$\vb{r}$ approach, the NAC expressed in Eq. (4) is the crucial
term determining the carrier dynamics. Accordingly, in the NAMD\_$\vb{k}$
approach, the \textit{e-ph} coupling $H_{{m\vb{k}'}, n\vb{k}}^{ep} =
\mel{\psi_{m\vb{k}'}}{\Delta V}{\psi_{n\vb{k}}}$ between states
$\psi_{m\vb{k}'}$ and $\psi_{n\vb{k}}$ plays the key role. In Figure~\ref{fig2}
(e-f) we plot the averaged $H_{{m\vb{k}'}, n\vb{k}}^{ep}$, where the x and y
axes represent the energy of $\psi_{m\vb{k}'}$ and $\psi_{n\vb{k}}$ (labeled as
$E_{m\vb{k}'}$ and $E_{n\vb{k}}$, respectively). In Figure~\ref{fig2} (e),
where $E_{m\vb{k}'}$ and $E_{n\vb{k}}$ range within [0.0, 1.5]~eV, the largest
$H_{{m\vb{k}'}, {n\vb{k}}}^{ep}$ can be roughly fitted by two lines, which are
expressed as $\vert E_{m\vb{k}'} - E_{n\vb{k}} \vert = \SI{0.2}{eV}$,
suggesting the coupling between two electronic states is the largest when the
state energy difference is around \SI{0.2}{eV}. Thus, when $E_\text{ini} >
\SI{0.2}{eV}$, the hot electron prefers to relax to an electronic state
\SI{0.2}{eV} lower in energy, which explains the quantized character with an
energy difference of \SI{0.2}{eV} observed in the fast relaxation process shown
in Figure~\ref{fig1} (d), suggesting that the hot electron relaxation is
strongly coupled to phonons with energy around \SI{0.2}{eV}.  When
$E_\text{ini} < \SI{0.2}{eV}$, as indicated by the square marked with the white
dashed lines in Figure~\ref{fig2} (f), the couplings between $E_{m\vb{k}'}$ and
$E_{n\vb{k}}$ are much smaller. The matrix elements close to the diagonal line,
where $\vert E_{m\vb{k}'} - E_{n\vb{k}} \vert$ is very small, play a crucial
role. This result implies that in this case, the coupling to the phonons with
small energies is essential.

\begin{figure}[ht]
\centering
\includegraphics[width=1.0\textwidth]{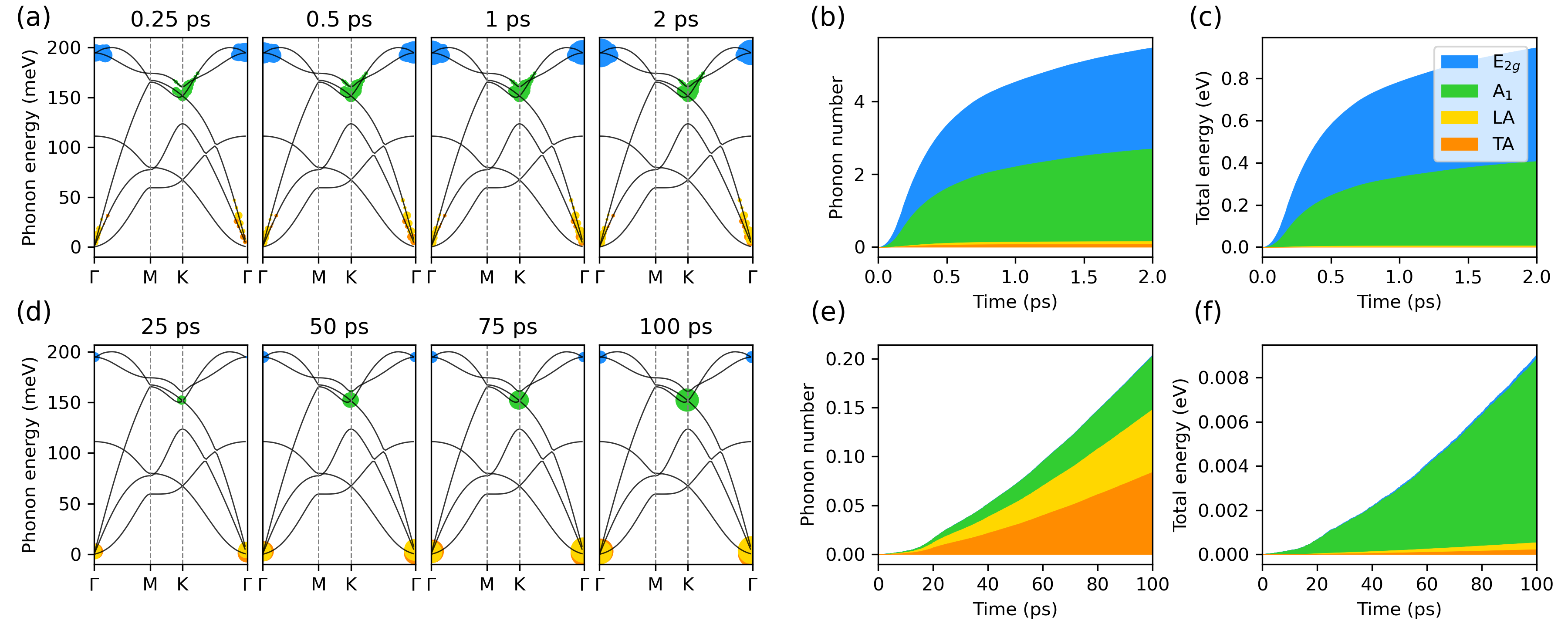}
\caption{
  Time-dependent phonon excitation during the hot electron relaxation dynamics.
  (a) Snapshots of phonon excitation dynamics at
  \SIlist[list-units=single]{0.25;0.5;1.0;2.0}{ps} with $E_\text{ini} =
  \SI{1.0}{eV}$ Time evolution of (b) the excited phonon number and (c) the
  total phonon energy during the hot electron relaxation with $E_\text{ini} =
  \SI{1.0}{eV}$. (d) Snapshots of the phonon excitaion dynamics at
  \SIlist[list-units=single]{25;50;75;100}{ps} with $E_\text{ini} =
  \SI{0.1}{eV}$. Time evolution of (e) the excited phonon number and (f) the
  total phonon energy during the hot electron relaxation with $E_\text{ini} =
  \SI{0.1}{eV}$.
  }\label{fig3}
\end{figure}

During the hot electron relaxation, the energy of the electrons transfers to
the phonons through the \textit{e-ph} coupling.  Figure~\ref{fig3} shows the
phonon excitation dynamics along with the hot electron relaxation.
Figure~\ref{fig3} (a) shows the four snapshots of phonon excitation within
\SI{2}{ps} with $E_\text{ini} = \SI{1.0}{eV}$. It can be seen that within the
hot electron relaxation process, only the optical modes $A_1$ and $E_{2g}$,
which belong to the LO and TO branches, are notably excited. In addition, there
is a minor excitation for the LA and TA modes.  Figure~\ref{fig3} (b-c) shows
the time-dependent phonon population and energies of these four different
phonon modes. It can be seen that the excitations of the optical $A_1$ and
$E_{2g}$ modes with $\hbar \omega_{\scriptscriptstyle A_1} = \SI{0.16}{eV}$ and
$\hbar\omega_{\scriptscriptstyle E_{2g}} = \SI{0.19}{eV}$ are dominant. The
excitation of $A_1$ is responsible for the intervalley electron scattering, and
both $A_1$ and $E_{2g}$ contribute to the quantized character in the energy
relaxation process.  Together, they define the critical energy threshold around
\SI{0.2}{eV}. Figure~\ref{fig3} (d-f) presents the time-dependent phonon
excitation with $E_\text{ini} = \SI{0.1}{eV}$.  In this case, the LA and TA
phonons around the $\Gamma$ point excitation are dominant, as shown in
Figure~\ref{fig3} (d) and (e). Due to the thermal energy smearing of the
electronic states (see more details in the Supplementary Materials), the
optical phonon mode $A_1$ also has a minor contribution. Since its energy is
much higher than the LA and TA phonon energies, its contribution to the excited
phonon energy is still dominant.  Since the energies and momenta of the LA and
TA phonons are both very small, the energy and valley dynamics are much slower.
The phonon excitation results explain the distinct hot electron relaxation
behavior when $E_\text{ini}$ is above or below the energy threshold of
\SI{0.2}{eV}.

\begin{figure}[ht]
\centering
\includegraphics[width=0.8\textwidth]{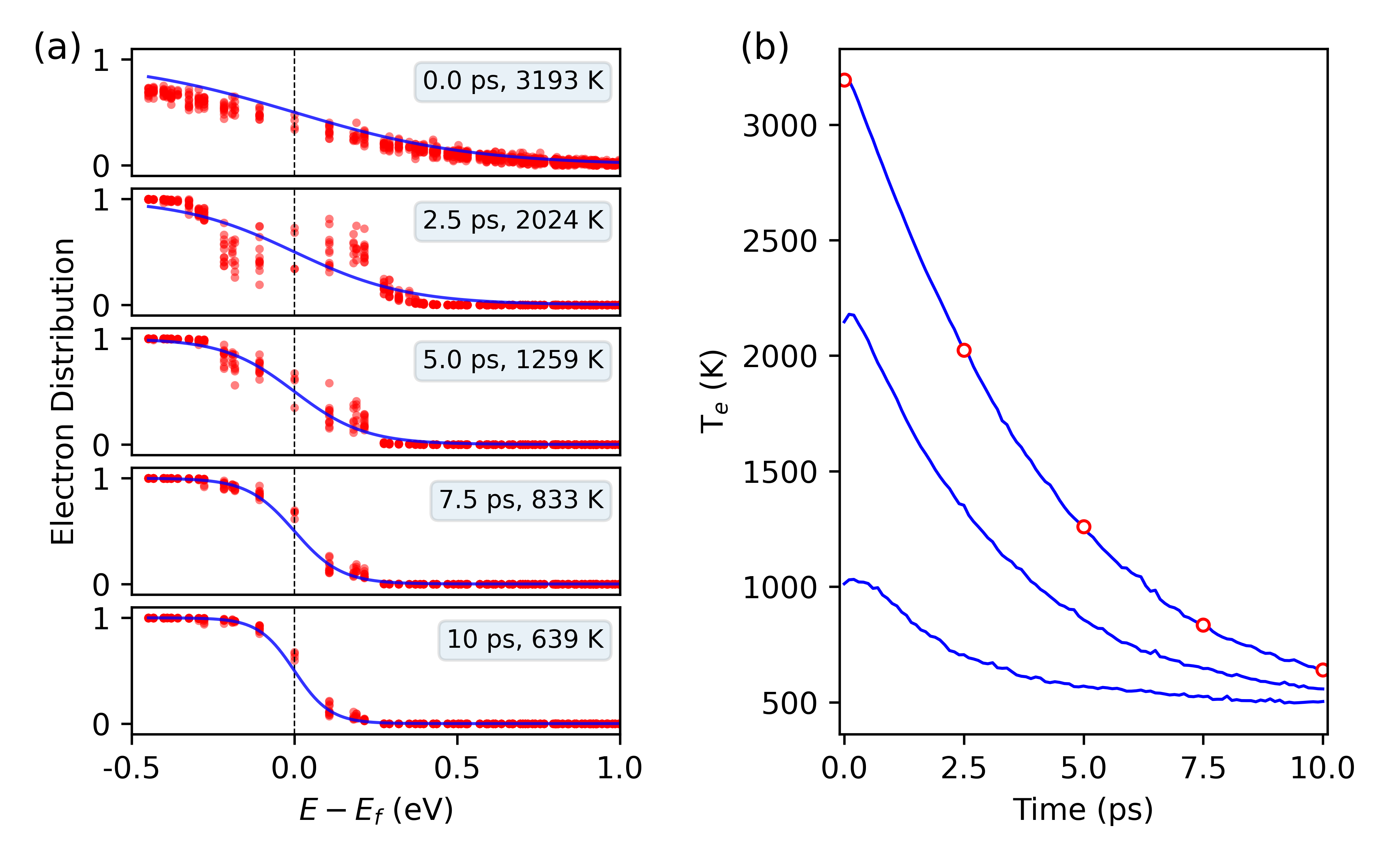}
\caption{
  (a) Snapshots of electron distribution at \SIlist[list-units=single]{0; 2.5;
  5.0; 7.5; 10}{ps} in multi-electron simulation with initial temperature of
  \SI{3000}{\kelvin}. (b) Evolution of electron temperature for initial
  temperature of \SIlist[list-units=single]{1000; 2000; 3000}{\kelvin}. The
  five panels in (a) correspond to the five small red circles in (b).
  }\label{fig4}
\end{figure}

Finally, we study the multi- hot electron relaxation by simulating the electron
temperature ($T_e$) decrease. In the TR-ARPES measurements, after
photoexcitation, the hot electrons will reach equilibrium with a certain
temperature through electron-electron (\textit{e-e}) scattering, and then relax
to a lower temperature through \textit{e-ph} coupling. Figure~\ref{fig4} (a)
shows 5 snapshots in the $T_e$ relaxation with initial electron temperature
$T_e^\text{ini} = \SI{3193}{\kelvin}$. In this case $T_e$ decreases to
\SI{639}{\kelvin} at \SI{10}{ps}.  Figure~\ref{fig4} (b) shows the
time-dependent relaxation dynamics with $T_e^\text{ini} =
\SIlist[list-units=single]{3193;2200;1060}{\kelvin}$. For all three cases,
$T_e$ converges to around \SI{500}{\kelvin} at \SI{10}{ps}. The relaxation from
\SI{500}{\kelvin} to lower temperature is very slow.

The simulation by NAMD\_$\vb{k}$ can explain most important experimental
results. Different experimental groups reported that when $E_\text{ini} <
\SI{0.2}{eV}$, slow decay via acoustic phonons plays a role.\cite{RN582, RN581,
RN580, RN578, RN590} while when $E_\text{ini} > \SI{0.2}{eV}$, the hot
electrons can efficiently scatter with the optical phonons.\cite{RN241, RN489,
RN582, RN590} The timescale is of the same magnitude as our results.
Especially, the strong coupling with the $A_1$ and $E_{2g}$ modes and the
quantized energy-loss are in excellent agreement with the report by Na et
al.\cite{RN489}. Our work also agrees with the theoretical study by Bernardi
and co-workers based on the quasi-classical time-dependent Boltzmann transport
equation (rt-BTE).\cite{RN567} It can be noted that hot electron relaxation
mainly excites $A_1$, $E_{2g}$, LA and TA. All of these phonons are in-plane
modes. The out-of-plane phonon modes are not directly excited, and thus we
propose the the buckling of graphene requires a phonon-phonon interaction,
which can be studied by the rt-BTE method.\cite{RN567}

Comparing with NAMD\_$\vb{r}$, the NAMD\_$\vb{k}$ method has a significant
advantage treating solid state systems. First, only the \textit{e-ph} coupling
based on the calculation with the unit cell is required for NAMD\_$\vb{k}$. By
contrast, for NAMD\_$\vb{r}$, the AIMD simulation and the electronic structure
calculations for each time step along the AIMD trajectory using a supercell are
required. The computational cost is significantly reduced by NAMD\_$\vb{k}$.
Second, the $\ket{\psi_{n\vb{k}}(\vb{r};\vb{R}_0)}$ basis sets used here can be
understood as diabatic basis sets. Each electronic state has a clear notion of
band index and momentum. We do not need to re-order the electronic states when
they cross each other, thus the well-known state crossing problem in surface
hopping can be avoided.\cite{RN121, RN122, RN127, RN258, RN252} Third, in the
NAMD\_$\vb{k}$ method, the \textit{e-ph} coupling can be understood in the time
domain. Especially, the time-dependent phonon excitation induced by electron
relaxation can be achieved, which is helpful to understand the photo-induced
lattice structure distortion, which can not be achieved by the NAMD\_$\vb{r}$
method using CPA, where the lattice distortion is pre-determined by the AIMD
simulation. \cite{RN222, RN515}  It also provides a dynamical picture of the
energy transport between the electron and the phonon subsystems at the
\textit{ab initio} level beyond the semi-classical two-temperature
model.\cite{RN603, RN604} Finally, since the \textit{e-ph} coupling considering
the SOC effects, as well as the exciton-phonon coupling elements can be
calculated based on previous studies,\cite{RN497, RN570} the NAMD\_$\vb{k}$
method can be easily extended for the investigation of SOC promoted spin
dynamics and exciton dynamics using GW + real-time Bethe-Salpeter equation
(GW+rtBSE) framework. The major approximation used in the NAMD\_$\vb{k}$ method
is the harmonic approximation. It will be very important and interesting to
consider anharmonic effects in this approach in the future.

\section{Summary}\label{sec4}

The newly developed NAMD\_$\vb{k}$ approach breaks the bottleneck of NAMD
application to carrier dynamics in momentum space. Based on the harmonic
approximation, the \textit{e-ph} coupling matrix can be used to calculate the
coupling between electronic states with different momenta, and the phonon
excitation dynamics can be simulated simultaneously.   The NAMD\_$\vb{k}$
approach is applied to study the hot carrier dynamics in graphene. It is found
that the optical phonon mode defines an energy threshold which separates the
hot electron carrier dynamics into fast and slow regions with lifetimes of ps
and ns, respectively. We propose the NAMD\_$\vb{k}$ approach to be an
indispensable tool in exploring excited carrier dynamics in the momentum space,
paving a new pathway to the theoretical design of devices and materials for
optoelectronics, valleytronics and solar energy conversion.

\bmhead{Acknowledgments}
J.Z.\ acknowledges the support of the National Natural Science Foundation of
China (NSFC), grant nos.\ 12125408, 11974322. Q.Z.\ acknowledges the support of
the National Natural Science Foundation of China (NSFC), grant nos.\ 12174363.
O.V.P.\ acknowledges funding of the US National Science Foundation, grant no.\
CHE-2154367.  YL.S. contributed to this work before March 2022.  Calculations
were performed Hefei Advanced Computing Center, and Supercomputing Center at
USTC.

\bibliography{ref}

\end{document}


\title[Article]{
\begin{center}
\textbf{Supplemental Materials} \\
for \\[12pt]
\textit{Ab initio} Real-Time Quantum Dynamics of Charge Carriers in Momentum Space
\end{center}
}

\author[1]{\fnm{Zhenfa} \sur{Zheng}}
\author*[1,2,3]{\fnm{Yongliang} \sur{Shi}} \email{sylcliff@xjtu.edu.cn}
\author[4]{\fnm{Jin-jian} \sur{Zhou}}
\author[5]{\fnm{Oleg V.} \sur{Prezhdo}}
\author*[1]{\fnm{Qijing} \sur{Zheng}} \email{zqj@ustc.edu.cn}
\author*[1,6]{\fnm{Jin} \sur{Zhao}} \email{zhaojin@ustc.edu.cn}

\affil*[1]{
    \orgdiv{Department of Physics, ICQD/Hefei National Laboratory for Physical
            Sciences at Microscale, and CAS Key Laboratory of Strongly-Coupled
            Quantum Matter Physics},
    \orgname{University of Science and Technology of China},
    \orgaddress{\city{Hefei}, \postcode{230026}, \state{Anhui}, \country{China}}
    }

\affil*[2]{
    \orgdiv{Center for Spintonics and Quantum Systerms, State Key Laboratory
            for Mechanical Behavior of Materials, School of Materials Science
            and Engineering},
    \orgname{Xi'an Jiaotong University},
    \orgaddress{\city{Xi'an}, \postcode{710049}, \state{Shanxi}, \country{China}}
    }

\affil*[3]{
    \orgdiv{China 2 State Key Laboratory of Surface Physics and Department of Physics},
    \orgname{Fudan University},
    \orgaddress{\city{Shanghai}, \postcode{200433}, \country{China}}
    }

\affil[4]{
    \orgdiv{School of Physics},
    \orgname{Beijing Institute of Technology},
    \orgaddress{\city{Beijing}, \postcode{100081}, \country{China}}
    }

\affil[5]{
    \orgdiv{Departments of Chemistry, Physics, and Astronomy},
    \orgname{University of Southern California},
    \orgaddress{\city{Los Angeles}, \postcode{90089}, \state{California}, \country{USA}}
    }

\affil*[6]{
    \orgdiv{Department of Physics and Astronomy},
    \orgname{University of Pittsburgh},
    \orgaddress{\city{Pittsburgh}, \postcode{15260}, \state{Pennsylvania}, \country{USA}}
    }

\maketitle

\section{Zero nonadiabatic coupling for Bloch states with different momenta}

In the real-space \textit{ab initio} nonadiabatic molecular dynamics (NAMD)
approach (NAMD\_$\vb{r}$), the charge carrier dynamics is mainly determined by
the nonadiabatic coupling (NAC), which can be written as:
\begin{align}\label{eq:NAC}
    d_{m\vb{k}', n\vb{k}}
    &= \mel{\psi_{m\vb{k}'}}{\dv t}{\psi_{n\vb{k}}}
    \nonumber \\
    &\approx \frac{
       \braket{\psi_{m\vb{k}'}(t)}{\psi_{n\vb{k}}(t+\Delta t)} -
       \braket{\psi_{m\vb{k}'}(t+\Delta t)}{\psi_{n\vb{k}}(t)}
       }{2\Delta t}.
\end{align}
For periodic systems, Bloch's theorem states that electronic state takes the
form of a Bloch wave, i.e.
\begin{equation}\label{eq:Bloch}
    \psi_{n\vb{k}}(\vb{r})
    = N_p^{-1/2} e^{i\vb{k}\vdot\vb{r}} u_{n\vb{k}}(\vb{r}),
\end{equation}
where $\vb{r}$ is the position of the electron, $N_p$ is the number of unit
cells, and $u_{n\vb{k}}(\vb{r})$ is a cell-periodic function, which satisfies
\begin{equation}
    u_{n\vb{k}}(\vb{r} + \vb{L}_p) = u_{n\vb{k}}(\vb{r}),
\end{equation}
where $\vb{L}_p$ is the lattice vector of the $p$-th unit cell. Note that the
Bloch wave $\psi_{n\vb{k}}$ is normalized in the Born-von K\'arm\'an (BvK)
supercell, while the periodic part $u_{n\vb{k}}(\vb{r})$ is normalized in the
unit cell. The overlap or inner product of two Bloch waves at different times
becomes
\begin{align}
   \braket{\psi_{m\vb{k}'}(t')}{\psi_{n\vb{k}}(t)}
   &= N_p^{-1} \int e^{-i(\vb{k}' - \vb{k})\vdot\vb{r}}
      u_{m\vb{k}'}^*(\vb{r})\,u_{n\vb{k}}(\vb{r}) \,\mathrm{d}\vb{r}
   \nonumber \\
   &= N_p^{-1} \sum_p \int\limits_{p\text{-cell}}
      e^{-i(\vb{k}' - \vb{k}) \vdot (\vb{r} + \vb{L}_p)}
      u_{m\vb{k}'}^*(\vb{r} + \vb{L}_p)\,u_{n\vb{k}}(\vb{r} + \vb{L}_p)
      \,\mathrm{d}\vb{r}
   \nonumber \\
   &= N_p^{-1} \sum_p e^{-i(\vb{k}' - \vb{k}) \vdot \vb{L}_p}
      \int_{\textmd{uc}} e^{-i(\vb{k}' - \vb{k})\vdot\vb{r}}
      u_{m\vb{k}'}^*(\vb{r})\,u_{n\vb{k}}(\vb{r}) \,\mathrm{d}\vb{r}
   \nonumber \\
   &= \delta_{\vb{k},\vb{k}'}
      \int_{\textmd{uc}} e^{-i(\vb{k}' - \vb{k})\vdot\vb{r}}
      u_{m\vb{k}'}^*(\vb{r};t')\,u_{n\vb{k}}(\vb{r};t) \,\mathrm{d}\vb{r},
      \label{eq:innprd}
\end{align}
where the subscripts ``$p$-cell'' and ``$\textmd{uc}$'' indicate that the
integral is carried out within the $p$-th unit cell and the first unit cell,
respectively.  From the third line to the last line, we have used
\begin{equation}
 N_p^{-1} \sum_p e^{-i(\vb{k}' - \vb{k}) \vdot \vb{L}_p} 
 = \delta_{\vb{k},\vb{k}'}.
\end{equation}
From Eq.~\eqref{eq:innprd}, one can see that the inner product of two Bloch
states at different times is zero for $\vb{k}' \neq \vb{k}$, thus the relating
NAC is also zero. Apparently, this excludes the possibility of using a unit
cell to perform NAMD calculations were one to include different Bloch states.
To circumvent this problem, in the NAMD\_$\vb{r}$ approach, by introducing a
large enough supercell, one can effectively fold the electronic states of
different momenta $\vb{k}$ to the Brillouin Zone center ($\Gamma$-point).
Moreover, a large supercell can also accommodate phonon modes  of different
wavevectors $\vb{q}$, which can compensate the momentum difference  needed for
the crystal momentum conservation, resulting in a non-zero overlap  and hence a
non-zero NAC between the states.

\section{Solving the Time-dependent Kohn-Sham (TDKS) equation}

For quantum dynamics of excited carrier, we solve the time-dependent
Shr\"{o}dinger equation (TDSE)
\begin{equation}\label{eq:TDSE}
  i\hbar\pdv{t} \ket{\Psi(\vb{r};\vb{R}(t))}
  = \hat{H}^{el}(\vb{r};\vb{R}(t)) \ket{\Psi(\vb{r};\vb{R}(t))}.
\end{equation}
Here, both electronic Hamiltonian $\hat{H}^{el}(\vb{r};\vb{R}(t))$ and
electronic wavefunction $\ket{\Psi(\vb{r};\vb{R}(t))}$ depend on time through
nuclear trajectory $\vb{R}(t)$.

In the nonadiabatic molecular dynamics simulation (NAMD) approach in momentum
space (NAMD\_$\vb{k}$), starting with the equilibrium atomic structure
$\vb{R}_0$, the electronic Hamiltonian can be separated into two parts
\begin{equation}\label{eq:Ham}
  \hat{H}^{el}(\vb{r};\vb{R}(t))
  = \hat{H}^0(\vb{r};\vb{R}_0) + \Delta\hat{H}(\vb{r};\vb{R}(t)),
\end{equation}
where the first part $\hat{H}^0(\vb{r};\vb{R}_0)$ is the Hamiltonian at
equilibrium positions $\vb{R}_0$, which is time-independent. The second part
$\Delta\hat{H}$ is the variation of Hamiltonian, which is induced by nuclear
displacements $\Delta\vb{R}(t)=\vb{R}(t)-\vb{R}_0$. As the nuclear coordinates
only appear in the potential term $V(\vb{r};\vb{R})$ in the electronic
Hamiltonian $\hat{H}^{el}=\hat{T}+\hat{V}$ , therefore $\Delta\hat{H}$ can be
also written as
\begin{equation}\label{eq:dH}
  \Delta\hat{H}(\vb{r};\vb{R}(t))
  = \Delta V(\vb{r};\vb{R}(t))
  = V(\vb{r};\vb{R}(t)) - V(\vb{r};\vb{R}_0).
\end{equation}
Then, we expand the electronic wavefunction by the eigenstates of equilibrium
structure Hamiltonian $\hat{H}^0(\vb{r};\vb{R}_0)$
\begin{equation}\label{eq:WavExpand}
  \ket{\Psi(\vb{r};\vb{R}(t))}
  = \sum_{n\vb{k}} c_{n\vb{k}}(t) \ket{\psi_{n\vb{k}}(\vb{r};\vb{R}_0)},
\end{equation}
where $\vb{k}$ represents the crystal momentum and $n$ is the band index. The
basis set $\qty\big{\ket{\psi_{n\vb{k}}(\vb{r};\vb{R}_0)}}$ are calculated in
the framework of Kohn-Sham (KS) density functional theory (DFT), and satisfy
\begin{equation}\label{eq:EquiSE}
  \hat{H}^0(\vb{r};\vb{R}_0) \ket{\psi_{n\vb{k}}}
  = \epsilon_{n\vb{k}} \ket{\psi_{n\vb{k}}},
\end{equation}
where $\epsilon_{n\vb{k}}$ is eigen-energy of KS state $\ket{\psi_{n\vb{k}}}$.
Here, it should be emphasized that the basis wavefunctions
$\qty\big{\ket{\psi_{n\vb{k}}}}$ are \emph{time-independent} while the
expanding coefficients $c_{n\vb{k}}(t)$ are \emph{time-dependent}.

By substituting Eq.~\eqref{eq:Ham}-\eqref{eq:EquiSE} into Eq.~\eqref{eq:TDSE}
and multiplying by $\bra{\psi_{m\vb{k}'}}$ on both sides of equation, we get
the time-dependent equation for the expanding coefficient
\begin{equation}\label{eq:TDCoef}
  i\hbar\dv{t}c_{m\vb{k'}}(t) = \sum_{n\vb{k}}
    \left(H^0_{m\vb{k}',n\vb{k}} + H^{ep}_{m\vb{k}',n\vb{k}}\right) c_{n\vb{k}}(t),
\end{equation}
where $H^0_{m\vb{k}',n\vb{k}}$ is the matrix element of $\hat{H}^0$, which is
apparently diagonal
\begin{equation}\label{eq:H0}
  H^0_{m\vb{k}',n\vb{k}} = \mel{\psi_{m\vb{k}'}}{\hat{H}^0}{\psi_{n\vb{k}}}
           = \epsilon_{n\vb{k}}\delta_{mn,\vb{k}'\vb{k}},
\end{equation}
and $H^{ep}_{m\vb{k}',n\vb{k}}$ is the electron-phonon (\textit{e-ph}) coupling
matix element
\begin{equation}\label{eq:Hep0}
  H^{ep}_{m\vb{k}',n\vb{k}} = \mel{\psi_{m\vb{k}'}}{\Delta V}{\psi_{n\vb{k}}}.
\end{equation}

To calculate the \textit{e-ph} term, we expand the potential energy
$V(\vb{r};\vb{R}(t))$ in terms of nuclear displacements $\Delta\vb{R}(t)$. The
potential to first order in displacements is
\begin{equation}\label{eq:dV0}
  V(\vb{r};\vb{R}(t))= V(\vb{r};\vb{R}_0) + \sum_{p\kappa}
  \eval{\pdv{V(\vb{r};\vb{R})}{\vb{R}_{p\kappa}}}_{\vb{R}=\vb{R}_0} \vdot
  \Delta\vb{R}_{p\kappa}(t).
\end{equation}
where the position of the nucleus $\kappa$ in the $p$-th unit cell is denoted
as $\vb{R}_{p\kappa}=\vb{L}_p+\vb{\tau}_{\kappa}$, and $\vb{L}_p$ is the cell
vector.  In practice, it is convenient to decompose nuclear displacements
$\Delta\vb{R}(t)$ into normal modes
\begin{equation}
  \Delta\vb{R}_{p\kappa}(t) = \sqrt{\frac{M_0}{N_pM_\kappa}}
  \sum_{\vb{q}\nu}
    e^{i\vb{q}\vdot\vb{L}_p} \vb{e}_{\kappa\nu}(\vb{q}) Q_{\vb{q}\nu}(t),
\end{equation}
where $M_0$ is an arbitrary reference mass introduced to ensure that both sides
of the equation have the dimension of length. Typically, $M_0$ is chosen to be
the proton mass. $\vb{e}_{\kappa\nu}(\vb{q})$ is the polarization of the
vibration wave corresponding to the wave vector $\vb{q}$ and mode $\nu$.
$Q_{\vb{q}\nu}(t)$ is the normal mode coordinate of the corresponding mode.
According to Eq.~\eqref{eq:dH},
\begin{equation}\label{eq:dV1}
    \Delta V(\vb{r};\vb{R}(t)) = \sqrt{\frac{1}{N_p}} \sum_{\vb{q}\nu} \sum_{p\kappa}
        \sqrt{\frac{M_0}{M_{\kappa}}} e^{i\vb{q}\vdot\vb{L}_p}
        \eval{\pdv{V(\vb{r};\vb{R})}{\vb{R}_{p\kappa}}}_{\vb{R}=\vb{R}_0}
        \kern-9pt\vdot \vb{e}_{\kappa\nu}(\vb{q}) Q_{\vb{q}\nu}(t).
\end{equation}
Here, by defining
\begin{equation}
    {\partial}_{\vb{q}\kappa} v(\vb{r};\vb{R}_0)
    = \sum_{p} e^{-i\vb{q}\vdot(\vb{r}-\vb{L}_p)}
      \eval{\pdv{V(\vb{r};\vb{R})}{\vb{R}_{p\kappa}}}_{\vb{R}=\vb{R}_0},
\end{equation}
and
\begin{equation}
    {\Delta}_{\vb{q}\nu} v(\vb{r};\vb{R}_0)
    = \sum_{\kappa} (M_0/M_{\kappa})^{1/2}
      {\partial}_{\vb{q}\kappa}V(\vb{r};\vb{R}_0) \vdot
      \vb{e}_{\kappa\nu}(\vb{q}) l_{\vb{q}\nu},
\end{equation}
where $l_{\vb{q}\nu}$ is the ``zero-point'' displacement amplitude:
\begin{equation}
    l_{\vb{q}\nu} = \sqrt{\frac{\hbar}{2M_0\omega_{\vb{q}\nu}}}
\end{equation}
One can show that $\partial_{\vb{q}\kappa}v(\vb{r};\vb{R}_0)$ and
${\Delta}_{\vb{q}\nu}v(\vb{r};\vb{R}_0)$ are lattice-periodic functions. Then,
Eq.~\eqref{eq:dV1} can be written as
\begin{equation}\label{eq:dV4}
    \Delta V(\vb{r};\vb{R}(t))
    = N_p^{-1/2} \sum_{\vb{q}\nu} e^{i\vb{q}\vdot\vb{r}} {\Delta}_{\vb{q}\nu}
      v(\vb{r};\vb{R}_0) Q_{\vb{q}\nu}(t)l_{\vb{q}\nu}^{-1}.
\end{equation}
By combining Eq.~\eqref{eq:Hep0}, \eqref{eq:dV4} and \eqref{eq:Bloch}, we have
\begin{align}
    H_{{m\vb{k}'},{n\vb{k}}}^{ep}
    &= N_p^{-1/2} \sum_{\vb{q}\nu} N_p^{-1} \mel{u_{m\vb{k}'}}{
       e^{-i(\vb{k}'-\vb{k}-\vb{q})\vdot\vb{r}}
       \Delta_{\vb{q}\nu}v(\vb{r};\vb{R}_0) }{u_{n\vb{k}}}
       Q_{\vb{q}\nu}(t) l_{\vb{q}\nu}^{-1} \nonumber\\
    &= N_p^{-1/2} \sum_{\vb{q}\nu}
       \mel{u_{m\vb{k}'}}{\Delta_{\vb{q}\nu}v(\vb{r};\vb{R}_0)}{u_{n\vb{k}}}_{\textmd{uc}}
       \delta_{\vb{q},\vb{k}'\text{-}\vb{k}} Q_{\vb{q}\nu}(t) l_{\vb{q}\nu}^{-1} \nonumber\\
    &= N_p^{-1/2} \sum_{\nu}
       g_{mn\nu}(\vb{k},\vb{q})
       Q_{\vb{q}\nu}(t) l_{\vb{q}\nu}^{-1}\,\biggl\vert_{\vb{q}=\vb{k}'\text{-}\vb{k}},
    \label{eq:Hep1}
\end{align}
where the inner-product in the first line is an integral over the entire BvK
supercell.  Similar to the proof of Eq.~\eqref{eq:innprd}, one can show that
the integral can be converted to an integral over the unit cell, provided that
the momentum is conserved.  $g_{mn\nu}(\vb{k},\vb{q})$ is referred to as the
\textit{e-ph} matrix element
\begin{equation}
    g_{mn\nu}(\vb{k},\vb{q})
    = \mel{u_{m\vb{k}+\vb{q}}}{\Delta_{\vb{q}\nu}v(\vb{r};\vb{R}_0)}
      {u_{n\vb{k}}}_{\textmd{uc}}.
\end{equation}

\section{Surface Hopping}

The fewest-switches surface hopping (FSSH) method represents time-evolving
electron-nuclear system by an ensemble of trajectories, propagating under the
influence of deterministic (via TDSE) and stochastic [via surface hopping(SH)]
factors. Evolution of the system of interest is defined in a joint space
combining classical phase-space for nuclei and discrete quantum states for
electrons. The paths are constructed such that the FSSH probabilities for all
states at all times, averaged over the trajectory ensemble, are equal to the
corresponding probabilities obtained from the TDSE.  The latter are given by
diagonal elements of the density matrix:
\begin{equation}
   \rho_{m\vb{k}', n\vb{k}}(t) = c_{m\vb{k}'}^*(t) c_{n\vb{k}}(t). 
\end{equation}
The off-diagonal elements determine the probabilities of transitions between
electronic states.  For simplicity, we take the diagonal elements
$\rho_{n\vb{k}, n\vb{k}}$ as $\rho_{n\vb{k}}$ following.  If the system is in
state ${n\vb{k}}$ at time $t$, then the probability to leave this state at time
$t + \Delta t$ is:
\begin{equation}
   P_{n\vb{k}}(t; t + \Delta t)
   = \frac{\rho_{n\vb{k}}(t)-\rho_{n\vb{k}}(t + \Delta t)}{\rho_{n\vb{k}}(t)} 
   = - \frac{\int_{t}^{t + \Delta t} \dot{\rho}_{n\vb{k}}(t) dt}{\rho_{n\vb{k}}(t)}.
\end{equation}
From the definition of the density matrix, and from the TDSE, it follows that:
\begin{equation}
   \dot{\rho}_{n\vb{k}}(t)
   = \sum_{m\vb{k}'} - \frac{2}{\hbar}
     \Im\left(
     \rho_{m\vb{k}', n\vb{k}}^* H_{m\vb{k}', n\vb{k}}^{ep}
     \right).
\end{equation}
Splitting the resulting hopping probability into various channels, $m\vb{k}'$,
one obtains the probability of transition between the pair of states $n\vb{k}
\rightarrow m\vb{k}'$:
\begin{equation}
   P_{n\vb{k} \rightarrow m\vb{k}'}(t;t+\Delta t)
   = \frac{2}{\hbar} \frac{
     \int_{t}^{t + \Delta t} \Im\left(
     \rho_{m\vb{k}', n\vb{k}}^* H_{m\vb{k}', n\vb{k}}^{ep}
     \right) dt
     }{\rho_{n\vb{k}}(t)}.
     \label{eq:SHProp}
\end{equation}
Note that each transition from $n\vb{k}$ to $m\vb{k}'$ corresponds to one
\textit{e-ph} scattering process, where the energy and crystal momentum must be
conserved. However, Eq.~\eqref{eq:SHProp} can not guarantee the conservation
law. To explicitly introduce the conservation law into the probability
equation, recall that Fermi's golden rule gives the transition rate between two
states
\begin{align}
    \dv{t} \mathcal{P}_{n\vb{k} \rightarrow m\vb{k}'}(t)
    = &\frac{2\pi}{\hbar}\vert H^{ep}_{m\vb{k}', n\vb{k}} \vert^2 \times
      \nonumber \\
      &\left[
      \delta(\epsilon_{m\vb{k}'} - \epsilon_{n\vb{k}} - \hbar\omega_{\vb{q}\nu}) +
      \delta(\epsilon_{m\vb{k}'} - \epsilon_{n\vb{k}} + \hbar\omega_{\vb{q}\nu})
      \right].
\end{align}
As a result, the modified hopping probability equation changes to
\begin{equation}
    \tilde{P}_{n\vb{k} \rightarrow m\vb{k}'}(t;t+\Delta t)
    = \frac{2}{\hbar}
      \frac{
      \Im\left(
      \rho_{m\vb{k}', n\vb{k}}^*(t) \tilde{H}_{m\vb{k}', n\vb{k}}^{ep}
      \right) \Delta t,
      }{\rho_{n\vb{k}}(t)},
\end{equation}
where $\tilde{H}_{m\vb{k}', n\vb{k}}^{ep}$ is given by
\begin{align}
    \tilde{H}_{m\vb{k}', n\vb{k}}^{ep}
    = &\frac{2\pi}{N_p} \sum_{\nu} \vert g_{mn\nu}(\vb{k},\vb{q})\vert ^2
      (n_{\vb{q}\nu} + \frac{1}{2})
      \times \nonumber \\
     &\left[
      \delta(\epsilon_{m\vb{k}'} - \epsilon_{n\vb{k}} - \hbar\omega_{\vb{q}\nu}) +
      \delta(\epsilon_{m\vb{k}'} - \epsilon_{n\vb{k}} + \hbar\omega_{\vb{q}\nu})
      \right].
\end{align}
where the two $\delta$ function correspond to absorbing and emitting of phonon,
respectively. In practical implementation, however, Gaussian function with
certain broadening $\sigma$ is used to approximate the Dirac $\delta$-function
\begin{equation}
    \delta (E)
    \approx\frac{1}{\sigma \sqrt{2\pi}}
            e^{-\frac{E^2}{2\sigma^2}}.
\end{equation}
If the computed probability is negative, it is reset to zero. Thus, in general,
the FSSH assigns a probability for transition from the current electronic state
${n\vb{k}}$ to the new state $m\vb{k}'$, as:
\begin{align}
   g_{n\vb{k} \rightarrow m\vb{k}'}(t;t + \Delta t)
   &= \max\left[0, \tilde{P}_{n\vb{k} \rightarrow m\vb{k}'}(t;t + \Delta t)\right],\\
   g_{n\vb{k} \rightarrow n\vb{k}}(t; t + \Delta t)
   &= 1 - \sum_{m\vb{k}' \neq n\vb{k}}g_{n\vb{k} \rightarrow m\vb{k}'}(t; t + \Delta t).
\end{align}
In order to reflect the detailed balance condition, the hop rejection and
velocity rescaling of the standard FSSH are replaced in the FSSH-CPA by scaling
the transition probabilities $g_{n\vb{k} \rightarrow n\vb{k}}$ with the
Boltzmann factor:
\begin{gather}
    g_{n\vb{k} \rightarrow m\vb{k}'} 
    \quad\Rightarrow\quad
    g_{n\vb{k} \rightarrow m\vb{k}'} b_{n\vb{k} \rightarrow m\vb{k}'},
\\[6pt]
    b_{n\vb{k} \rightarrow m\vb{k}'} = 
    \begin{cases}
        \exp\left( - \frac{\epsilon_{m\vb{k}'} - \epsilon_{n\vb{k}}}{k_BT} \right)
        & \epsilon_{m\vb{k}'} > \epsilon_{n\vb{k}} \\
        1 & \epsilon_{m\vb{k}'} \leq \epsilon_{n\vb{k}}
    \end{cases}.
\end{gather}

\section{Test of k-grid Size and Energy Broadening}

\begin{figure}[ht]
    \centering
    \includegraphics[width=1.0\linewidth]{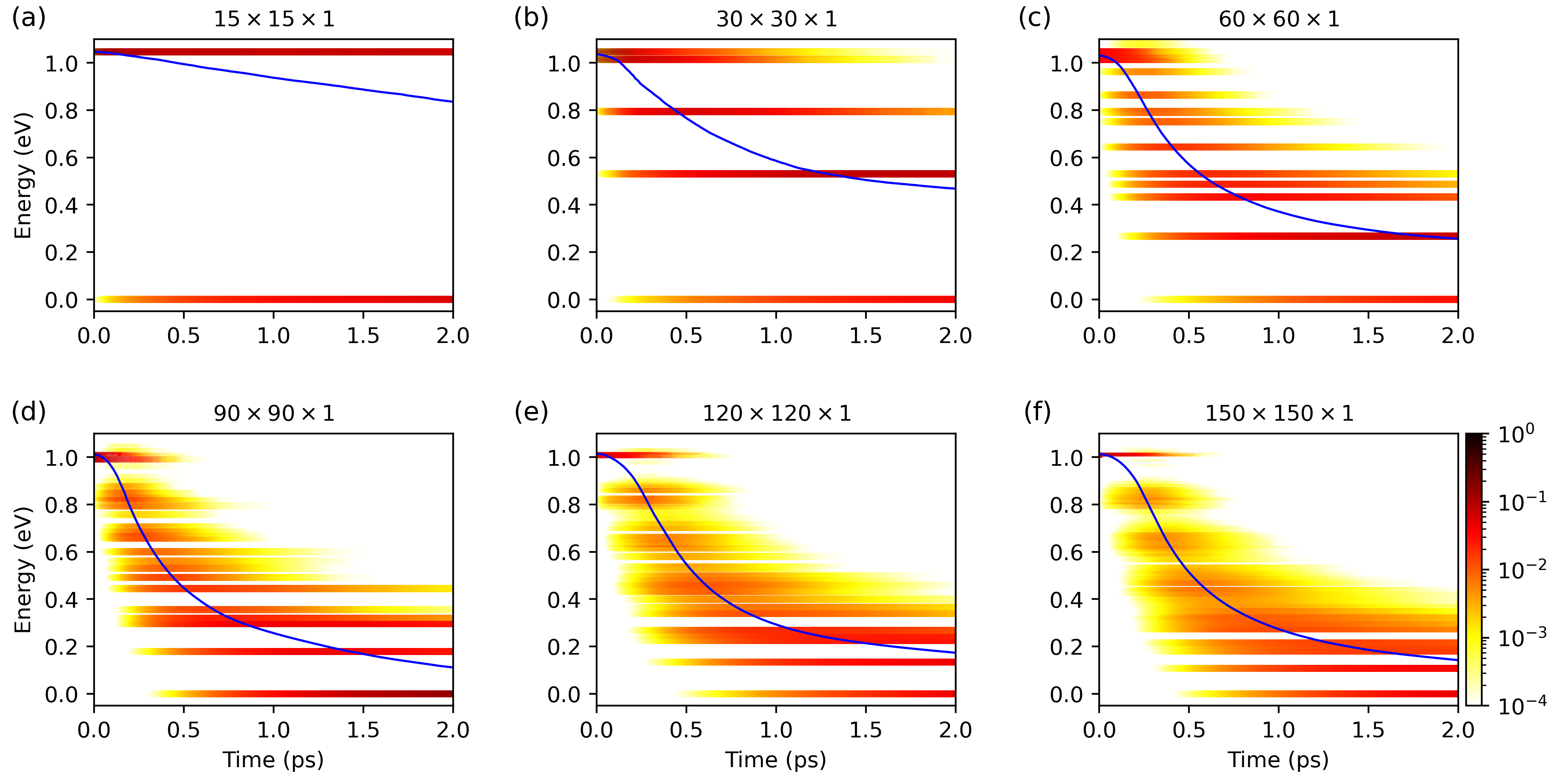}
    \caption{
      Hot electron dynamics ($E_{ini} = \SI{1.0}{eV}$) with $\vb{k}$-points
      grid of (a) $15\times15\times1$, (b) $30\times30\times1$, (c)
      $60\times60\times1$, (d) $90\times90\times1$, (e) $120\times120\times1$,
      and (f) $150\times150\times1$.  The color dots indicate the electron
      population in different states, and the blue line represents the averaged
      electron energy.
      }\label{fig:S1}
\end{figure}

In graphene, the Dirac conical band structure results in a very small density
of electronic states near the Fermi level ($E_f$), so a sufficiently dense
$\vb{k}$-grid is required to accurately model the hot electron relaxation
process. We have tested the simulation of hot electron relaxation with
different $\vb{k}$-grid sizes, as shown in Fig.~\ref{fig:S1}. It can be seen
that with the increase of $\vb{k}$-point density, the electron relaxation
process gradually converges. When the density of $\vb{k}$-points reaches
$90\times90\times1$, the change of the hot electron relaxation process becomes
negligible. However, it should be noted that as the electron energy gets closer
to the Dirac point, the density of its electronic states decreases, so a denser
$\vb{k}$-point is required. Hence, we use the $\vb{k}$-point grid of
$150\times150\times1$ in simulations for $E_\text{ini} =
\SIrange[range-phrase=-, range-units=single]{0.4}{1.5}{eV}$, and $\vb{k}$-point
grid of $450\times450\times1$ in simulations for $E_\text{ini} < \SI{0.3}{eV}$.

\begin{figure}[ht]
    \centering
    \includegraphics[width=1.0\linewidth]{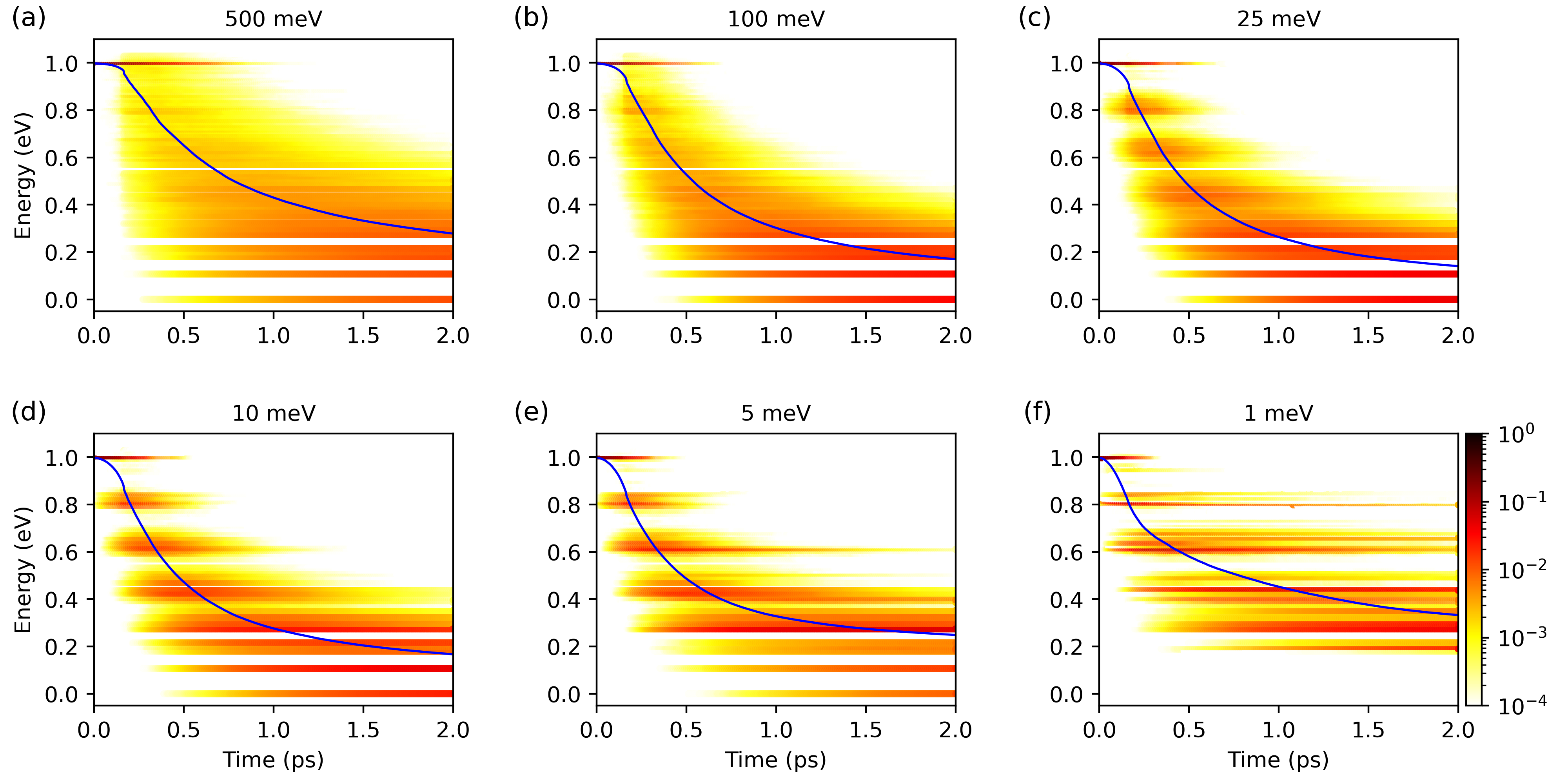}
    \caption{
      Hot electron dynamics ($E_\text{ini} = \SI{1.0}{eV}$) with energy
      broadening of (a) \SI{500}{meV}, (b) \SI{100}{meV}, (c) \SI{25}{meV}, (d)
      \SI{10}{meV}, (e) \SI{5}{meV}, and (f) \SI{1}{meV}.  The color dots
      indicate the electron population in different states, and the blue line
      represents the averaged electron energy.
      }\label{fig:S2}
\end{figure}

In the \textit{e-ph} coupling with energy conservation correct, to choose a
reasonable energy broadening, we have tried different $\sigma$ values to
simulate the hot electron dynamics with initial energy of \SI{1.0}{eV}
($E_\text{ini} = \SI{1.0}{eV}$). As shown in Fig.~\ref{fig:S2}, the electron
dynamics gets more and more elaborate as the energy broadening decreases 500 to
\SI{5}{meV}, while the relaxation time barely changes.  We can also find that
the relaxation is blocked at low energy region for energy broadening is smaller
than \SI{10}{meV}. That is due to the finite $\vb{k}$-grid, which causes energy
gap between different electronic states. To simulate the carrier relaxation in
energy bands with continuous $\vb{k}$-grid, usually a smaller $\sigma$ will
provide more accurate results if the $\vb{k}$-grid sampling is dense enough. In
this article we use $\sigma$ = 25 meV. To simulate the carrier dynamics across
a band gap,  the $\sigma$ value can be estimated by the thermal oscillation of
the electronic state at a certain temperature.   

\section{Comparison with NAMD simulation in real space (NAMD\_{r})}

\begin{figure}[ht]
    \centering
    \includegraphics[width=1.0\linewidth]{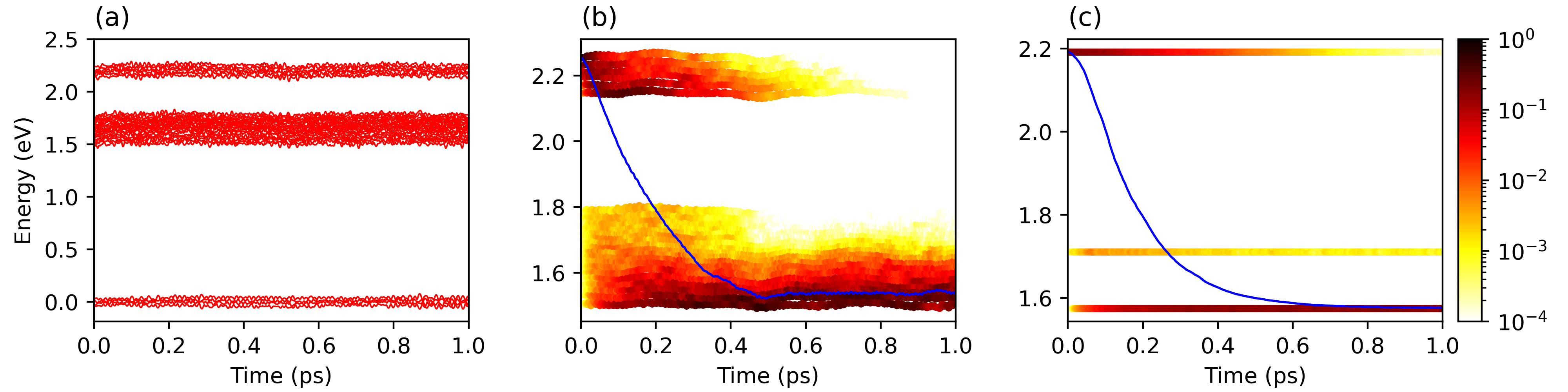}
    \caption{
      (a) Energy evolution of KS states in the NAMD\_$\vb{r}$ approach.  The
      hot electron relaxation simulation from \SI{2.2}{eV} to \SI{1.6}{eV}
      using (b) NAMD\_$\vb{r}$ method with $9\times9\times1$ supercell and (c)
      NAMD\_$\vb{k}$ method with $9\times9\times1$ $k$-grid.  The color dots
      indicate the electron population in different states, and the blue line
      represents the averaged electron energy.
      }\label{fig:S3}
\end{figure}

In order to verify the validity of the NAMD\_$\vb{k}$ method, we compare the
calculation results of the NAMD\_$\vb{k}$ with the NAMD\_$\vb{r}$ approach. In
the NAMD\_$\vb{r}$ approach, a supercell is required to sample the phonon
excitation with different $\vb{q}$ points.  Here we only simulate a
$9\times9\times1$ supercell due to the limitation of computational cost. Then
we compare the results with what NAMD\_$\vb{k}$ obtains using $9\times9\times1$
$\vb{k}$-grid. As the energy gaps are large for the $9\times9\times1$ supercell
or $\vb{k}$-grid close to the Fermi level, we choose to simulate the process of
hot electron evolving from \SI{2.2}{eV} to \SI{1.6}{eV}, where the energy gaps
are smaller relatively. As shown in Fig.~\ref{fig:S3}, the states in
NAMD\_$\vb{k}$ simulation degenerate to 3 energy levels because we use the
eigenstates of equilibrium structure Hamiltonian, and in fact the number of
states equals to the NAMD\_$\vb{r}$ simulation. Based on the NAMD\_$\vb{r}$
simulation on graphene with $9\times9\times1$ supercell, we estimate the
$\sigma$ value to be 120 meV from the thermal oscillation. The results obtained
by NAMD\_$\vb{r}$ and NAMD\_$\vb{k}$ agree with each other in substance. It
clearly proves the validity of the NAMD\_$\vb{k}$ approach.